\documentclass[12pt]{article}
\usepackage{amsmath}
\usepackage{graphicx}
\usepackage{enumerate}
\usepackage{natbib}
\usepackage{url} %

\usepackage{color}
\usepackage{verbatim}
\usepackage{booktabs}
\usepackage{textcomp}
\newcommand{\blind}{1}

\usepackage{amsmath,bm,nccmath,natbib,amsthm}
\usepackage{subcaption}
\usepackage{amsfonts}
\usepackage{longtable}
\usepackage{array}
\usepackage{float}
\usepackage{caption}
\usepackage{multirow}
\usepackage{xcolor}
\usepackage{comment}
\usepackage{bbm}
\usepackage{amssymb}
\usepackage{graphicx}
\usepackage{booktabs} 
\usepackage{tabularx}

\definecolor{myblue}{RGB}{0,0,255}
\definecolor{mygreen}{RGB}{44,85,17}
\definecolor{myred}{RGB}{85,17,44}

\newtheorem{theorem}{Theorem}

\newtheorem{definition}{Definition}\theoremstyle{definition}
\newtheorem{remark}{Remark}
\newcommand{\MName}{DiPH}
\newcommand{\diag}{\mathrm{diag}}
\newcommand{\vecf}{\mathrm{vec}}
\newcommand{\rank}{\mathrm{rank}}

\newcommand{\bo}{{\bf 1}}

\newcommand{\bS}{{\mathbb{S}}}
\newcommand{\bR}{{\mathbb{R}}}

\newcommand{\cL}{{\mathcal{L}}}
\newcommand{\cV}{W}

\newcommand{\cE}{{\mathcal{E}}}

\newcommand{\cl}{\bar{\cL}} %

\DeclareMathOperator{\tr}{tr}

\date{}

\begin{document}

\def\spacingset#1{\renewcommand{\baselinestretch}%
{#1}\small\normalsize} \spacingset{1}

\if1\blind
{
  \title{\bf Modeling Hypergraphs with Diversity and Heterogeneous Popularity}
 \author{Xianshi Yu\hspace{.2cm}\\
   Department of Statistics, University of Michigan\\
   and \\
   Ji Zhu\hspace{.2cm} \\
   Department of Statistics, University of Michigan}
  \maketitle  
} \fi

\if0\blind
{
  \bigskip
  \bigskip
  \bigskip
  \begin{center}
    {\LARGE\bf Modeling Hypergraphs with Diversity and Heterogeneous Popularity}
\end{center}
  \medskip
} \fi

\bigskip
\begin{abstract}
While relations among individuals make an important part of data with scientific and business interests, existing statistical modeling of relational data has mainly been focusing on dyadic relations, i.e., those between two individuals. This article addresses the less studied, though commonly encountered, polyadic relations that can involve more than two individuals. In particular, we propose a new latent space model for hypergraphs using determinantal point processes, which is driven by the diversity within hyperedges and each node's popularity. This model mechanism is in contrast to existing hypergraph models, which are predominantly driven by similarity rather than diversity. Additionally, the proposed model accommodates broad
types of hypergraphs, with no restriction on the cardinality and multiplicity of hyperedges, which previous models often have. 
Consistency and asymptotic normality of the maximum likelihood estimates of the model parameters have been established. 
The proof is challenging, owing to the special configuration of the parameter space. Further, we apply the projected accelerated gradient descent algorithm to obtain the parameter estimates, and we show its effectiveness in simulation studies. We also demonstrate an application of the proposed model on the \textit{What's Cooking} data 
and present the embedding of food ingredients learned from cooking recipes using the model.  
\end{abstract}
\vfill{}

\newpage{}
\spacingset{1.9} %
\section{Introduction}

Data that represent relations and interactions are ubiquitous in business, science, engineering and medicine. Driven by the needs of analyzing relational data, the field of network data analysis has seen rapid development in recent years. The scope of traditional network data analysis mainly focuses %
on dyadic relations. However, polyadic relations that involve more than two individuals are even more common in real-world interactions. For example, a collaboration in producing a movie (or an academic publication) usually involves more than two people, a protein complex often consists of more than two interacting proteins \citep{klamt2009hypergraphs}, and a supermarket transaction commonly includes more than two purchased items. In current practice, polyadic relations are often projected into a dyadic network before any analysis, which obviously causes substantial loss of information \citep{greening2015higher,karwa2016discussion,chodrow2020configuration}. Thus, effectively extracting information from polyadic relational data requires studying them directly.

\begin{figure}[h!]
\centering \includegraphics[width=\textwidth]{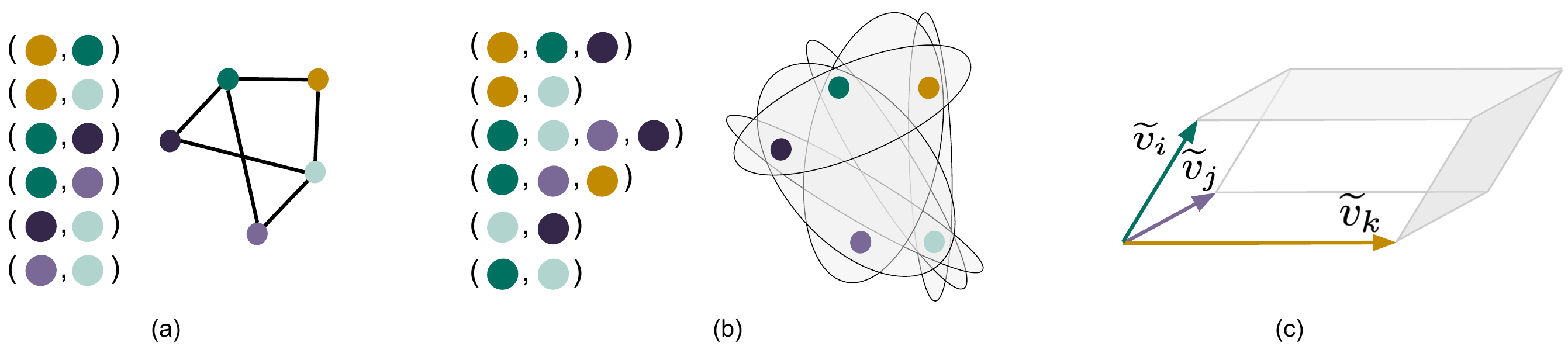}
\caption{(a) A traditional network and its edge list. (b) A hypergraph and its hyperedge list; each oval-shaped grey area represents a hyperedge comprising the nodes in it. (c) The parallelotope formed with $\widetilde{v}_i$, $\widetilde{v}_j$ and $\widetilde{v}_k$ (with $e=\{i,j,k\}$). Note the volume of the parallelotope is determined by the lengths and spread of $\widetilde{v}_i$, $\widetilde{v}_j$ and $\widetilde{v}_k$.}
\label{Fig::hypergraph_illu} 
\end{figure}

Relations that involve more than two individuals can be naturally represented using a hypergraph \citep{berge1970graphs}, which generalizes the traditional network. A hypergraph is defined by a set of nodes $\mathcal{V}$ and a (multi)set of hyperedges. Each hyperedge is a subset of $\mathcal{V}$, indicating existence of a relation among the nodes in the hyperedge. A subset of $\mathcal{V}$ of any size can be a hyperedge, in contrast to a traditional network which only allows edges to represent relations between exactly two nodes (see (a) and (b) in Figure \ref{Fig::hypergraph_illu}). Hypergraphs have been used in studies of chemical structures \citep{KONSTANTINOVA2001365}, relational database schemes \citep{fagin1983degrees}, and image processing \citep{ducournau2012reductive,yu2012adaptive,li2013contextual}, etc. More real-world data examples that can be naturally represented as hypergraphs can be found in \citet{benson2018simplicial}.

Many algorithms have been developed for hypergraphs, with two main foci. One is node clustering using generalization of graph cuts or modularity %
\citep{li2017inhomogoenous,veldt2020minimizing,kumar2020hypergraph,benson2020augmented,benson2021hypergraph}; the other is hypergraph embedding \citep{zhou2006learning,tu2018structural,maleki2021netvec}. These algorithms are heuristic and lack statistical models or principles.

Much of hypergraph modeling effort has been on a special type of hypergraph, the $k$-uniform hypergraph,
where all hyperedges have the same number of nodes $k$ \citep{ghoshdastidar2014consistency,ghoshdastidar2017uniform,chien2018community,kim2018stochastic,lyu2021latent,yuan2021high}. 
However, real-world hypergraphs are usually not $k$-uniform (e.g., number of engineers in an engineer project or number of items in a supermarket purchase). 

For non-uniform hypergraphs, the focus has been again on clustering \citep{ghoshdastidar2017consistency,ke2019community,chodrow2021generative,ng2021model}. %
There is also work on hereditary hypergraphs \citep{zhang2015exchangeable,lunagomez2017geometric}, where all subsets of a hyperedge are required to be hyperedges. 
Notable works not limited to $k$-uniform hypergraphs, hereditary hypergraphs or clustering include \citet{turnbull2019latent} and \citet{zhen2021community}, both of which assign latent positions to nodes. However, they do not allow for multiplicity of hyperedges. Multiple hyperedges are not uncommon in the real world observations (e.g., a group of engineers collaborating more than once, a set of products purchased together repeatedly); see \citet{benson2018simplicial} for many more examples.
In Table 1 in the Supplemental Material, we summarize all the hypergraph models mentioned above and compare their characteristics.

In this paper, we propose a new latent space model for non-uniform, non-hereditary hypergraphs which allows multiple hyperedges. 
Further, while most existing hypergraph models promote hyperedges among sets of similar nodes, where similarity is characterized by either latent positions or community labels, the proposed model prioritizes sets of nodes with high diversity in their latent positions. This is motivated by the observation that, in real-world scenarios, individuals that join together often have different characteristics and thus complement each other. For example, in collaborations, people with different specialties gather together and form a versatile and comprehensive team. This ``diversity'' is also often seen in human-made selections, where similar items are avoided to prevent redundancy.  Examples include the selection of products to purchase 
and the selection of tags to assign to an online item: while certain products or tags may be highly correlated, very similar ones tend to be avoided.  The proposed model also accounts for the variation in node popularity, as some nodes appear in hyperedges much more frequently than others.

The main contributions of this paper are summarized as follows. 
\begin{itemize}
\item We propose a new latent space model for hypergraphs, which is generative in hyperedges and thus allows multiple edges. To the best of our knowledge, it is the first hypergraph model that addresses diversity within hyperedges.  Moreover, it admits the variation in popularity among nodes.  The proposed model allows node sets of any cardinality to form hyperedges, and therefore, is able to model non-uniform hypergraphs. 

\item We establish the consistency as well as the asymptotic normality for the maximum likelihood estimate (MLE) of the parameters of the proposed model.  The challenge of this theoretical development lies in that the parameter space is a special manifold, which arises from the model configuration.  Further, while the proposed model can be considered as a determinantal point process (DPP) with special structures that enable compact parameterization, to the best of our knowledge, our theory is the first result that establishes the asymptotics regarding structured DPPs. 

\item We apply the accelerated projected gradient descent algorithm to obtain the MLE of the model parameters, which include each node's latent position and popularity parameter.  We demonstrate the efficacy of the algorithm in simulation studies. Moreover, we show that the estimated node embeddings can be used for further down stream tasks, such as node clustering.
We illustrate these utilities of the proposed model on a cooking recipe data example, where hyperedges are recipes and nodes are food ingredients. 
\end{itemize}

\noindent \textbf{\emph{Organization.}} The new hypergraph model is introduced in Section 2, along with discussions on the properties of the model. The algorithm for fitting the proposed model is also given in Section 2. In Section 3, we present theoretical results regarding the model parameter estimates, including their consistency and asymptotic normality. Sections 4 and 5 respectively present simulation studies and real-world data example results.%

\noindent \textbf{\emph{Notations.}} For a positive integer $n$, we use $[n]$ to denote $\{1, 2, \ldots, n\}$.  We use $\mathcal{V}$ to denote the set of all nodes and let $n_v:=\vert\mathcal{V}\vert$ be the number of nodes in $\mathcal{V}$.  We refer to nodes by indices in $[n_v]$.  Thus $\mathcal{V}$ and $[n_v]$ are used interchangeably.  The collection of observed hyperedges is denoted by $\mathcal{E}=\{e_1, e_2, \ldots, e_{n_e}\}$, where $e_\ell\subset \mathcal{V}$ (or $e_\ell\subset [n_v]$) for $\ell\in [n_e]$ and $n_e$ denotes the number of observed hyperedges.  Note that elements in $\mathcal{E}$ can be duplicated, as the same hyperedge can be observed repeatedly.
We use $\|\cdot\|_{2}$ to denote the $l_{2}$-norm of a vector and use $\|\cdot\|_{F}$ to denote the Frobenius norm of a matrix. We use $\det(\cdot)$ to denote the determinant of a matrix. For a subset $e\subset[n]$, and any $n$-by-$n$ matrix $M$, $M_e$ denotes the $\vert e \vert$-by-$\vert e \vert$ submatrix of $M$ corresponding to rows and columns indexed by $e$.  In addition, for any matrix $M$, $M_{i.}$ and $M_{.j}$ denote its $i$th row and $j$th column respectively.  Let $\bR^{n_{1}\times n_{2}}$ be the space of $n_1$-by-$n_2$ matrices with real entries and $\bS_n$ be the set of all $n$-by-$n$ symmetric matrices.  Further, $\mathbb{O}_{n}$ denotes the set of all $n$-by-$n$ orthogonal matrices, and $\mathbb{D}_{n}$ denotes the set of all $n$-by-$n$ diagonal matrices with diagonal entries in $\{-1,1\}$.  We use $\bo$ and ${\bf 0}$ to denote a vector of all ones and a vector of all zeros respectively.

\section{Model}

In this section, we propose a new generative latent space model for hypergraphs, which is driven by diversity within hyperedges and heterogeneous popularity among nodes.  Properties of the proposed model are discussed, and we also develop an algorithm for model fitting, which involves estimating nodes' latent positions and popularity parameters.

Recall that $\mathcal{V}$ denotes the set of nodes and $n_v=\vert\mathcal{V}\vert$ denotes the number of nodes in  $\mathcal{V}$. We associate each node $i$ ($i\in[n_v]$) with a latent position $v_{i}$ in $\bR^{d}$ $(0<d<n_v)$ and a popularity parameter $\alpha_{i}>0$. For a set of observed hyperedges $\{e_1, e_2, \ldots, e_{n_e}\}$, where $n_e$ denotes the number of hyperedges and each $e_\ell$ ($1\le \ell \le n_e$) is a subset of $[n_v]$, we assume that
\begin{equation}
e_\ell \overset{\text{{\tiny i.i.d.}}}{\sim} \mathcal{P} \text{ for } \ell=1,\cdots,n_e,
\label{definition::modeliid}
\end{equation}
where $\mathcal{P}$ denotes a distribution over all subsets of $[n_v]$, namely the power set of $[n_v]$, which has cardinality $2^{n_v}$ and contains all possible hyperedge configurations.

We use $E$ to denote a generic random hyperedge and $e$ to denote its realization. Given $v_i$'s and $\alpha_i$'s ($i\in [n_v]$), we model $\mathcal{P}$ as follows. Define vectors $\widetilde{v}_{i}$ ($i\in[n_v]$) by 
\begin{equation}
\widetilde{v}_i=(v_i,\sqrt{\alpha_{i}}w_i) \in\mathbb{R}^{d+n_v},
\label{Def::tildev}
\end{equation}
where $w_i$ is the $i$th standard basis vector of $\mathbb{R}^{n_v}$, i.e., the $i$th element is equal to 1 and all other elements are zero. Thus, $\widetilde{v}_{1}=(v_{1},\sqrt{\alpha_{1}},0,\cdots,0)$, $\widetilde{v}_{2}=(v_{2},0,\sqrt{\alpha_{2}},0,\cdots,0)$ and so on.  Given any hyperedge $e\subset[n_v]$, which is a subset of $\{1, \cdots, n_v\}$, we consider the parallelotope formed with vectors $\widetilde{v}_{i}$'s, $i\in e$ (see Figure \ref{Fig::hypergraph_illu} (c)) and define the distribution $\mathcal{P}$ of a random hyperedge $E$ by setting 
\begin{equation}
P(E=e) \propto \text{vol}^{2}\left(\left\{ \widetilde{v}_{i}, i\in e\right\} \right),
\label{Def::P}
\end{equation}
where $\text{vol}\left(\left\{ \widetilde{v}_{i}, i\in e\right\} \right)$ denotes the volume of the $\vert e \vert$-dimensional parallelotope formed by $\widetilde{v}_i$'s, $i\in e$. In particular, when $e$ has two (three) elements, $\{\widetilde{v}_{i}, i \in e \}$ form a parallelogram (parallelepiped), and $\text{vol}\left(\left\{ \widetilde{v}_{i}, i\in e\right\} \right)$ is its area (volume).  Let $L_{n_v\times n_v}:=(\widetilde{v}_{i}^\top \widetilde{v}_{j})_{i,j=1}^{n_v}$. It is well known (e.g., Theorem 7.5.1 in \cite{anderson2003introduction}) that the volume in (\ref{Def::P}) is given by 
\[
\text{vol}^{2}\left(\left\{ \widetilde{v}_{i}, i\in e\right\} \right)=\det(L_e),
\]
where $L_e$ denotes the submatrix of $L$ indexed by $e$, and $\det(\cdot)$ denotes the determinant of a matrix. Moreover, one can also show (see Supplemental Material D) that
\begin{equation}\label{equation::det}
\sum_{e\subset[n_v]}\det(L_e)=\det(L+I).
\end{equation}
Therefore, we have 
\begin{equation}
P(E=e)=\frac{\det(L_{e})}{\det(L+I)},
\label{definition::model}
\end{equation}
for any $e\subset[n_v]$. In addition, from (\ref{Def::tildev}) it is not difficult to see 
\begin{equation}
L_{n_v\times n_v}=(v_i^\top v_j)_{i,j=1}^{n_v}+\diag(\alpha), \text{ where }\alpha:=(\alpha_{1},\alpha_{2},\cdots,\alpha_{n_v}).
\label{definition::L}
\end{equation}
Overall, our proposed hypergraph model is defined by (\ref{definition::model}), (\ref{definition::L}), and (\ref{definition::modeliid}) together, and we name it as the Diversity and Popularity Hypergraph (\MName{}) model.  We write this model as
$H(v_{1},\cdots,v_{n_v},\alpha)$, or equivalently, $H(L)$ using (\ref{definition::L}).

To ensure identifiability, we constrain $\|v_i\|_2$ ($i\in[n_v]$) to be an (unknown) constant, i.e.
\begin{equation}
\|v_{i}\|_2=\|v_{i'}\|_2>0 ~\text{for any}~i,i'\in[n_v].
\label{constriant::v}
\end{equation}
The identifiability property is formally presented in Theorem \ref{Thm::identifiability}, which states that under the constraint \eqref{constriant::v}, $\alpha$ is identifiable and $v_{i}$'s are identifiable up to a universal orthogonal transformation and individual sign changes.  The proof of Theorem \ref{Thm::identifiability} is given in the Supplemental Material.

\begin{theorem}
\label{Thm::identifiability} 
For two \MName{} models $H(v_{1},\cdots,v_{n_v},\alpha)$ and $H(v_{1}',\cdots,v_{n_v}',\alpha')$, where $v_{i},v_{i}'\in\bR^{d}$ and at least one of $\{v_{i}\}_{i=1}^{n_v}$ and $\{v_{i}'\}_{i=1}^{n_v}$ spans $\bR^{d}$, if both satisfy (\ref{constriant::v}) and have $\alpha_{i},\alpha_{i}'>0$ for all $i\in[n_v]$, then the two hypergraph models are equivalent if and only if $\alpha=\alpha'$ and there exists an orthogonal matrix $O_{d\times d}$ such that $v_{i}'=\pm Ov_{i}$ for all $i\in[n_v]$, where the sign multiplied in front of $Ov_{i}$ may vary for different $i$'s. 

In addition, under the $H(L)$ parameterization, two \MName{} models $H(L)$ and $H(L')$ are equivalent if and only if there exists a diagonal matrix $D$ with diagonal entries in $\{-1,1\}$ such that $L'=DLD$.
\end{theorem}

Note that a given hyperedge $e\subset[n_v]$ will have a high probability $P(E=e)$ if the parallelotope spanned by $\{\widetilde{v}_{i}, i\in e\}$ has a comparatively large volume.  That, in turn, happens when $\|\widetilde{v}_{i}\|_{2}$'s ($i\in e$) are large and the directions of vectors  $\{\widetilde{v}_{i}, i\in e\}$ are well separated, i.e., as close to orthogonal as possible. Since $\widetilde{v}_{i}=(v_{i},\sqrt{\alpha_{i}}{w}_{i})$ and $\sqrt{\alpha_{i}}{w}_{i}$'s ($i\in[n_v]$) are pairwise orthogonal, the angles between $\widetilde{v}_{i}$'s are largely determined by $v_{i}$'s, and $\{\widetilde{v}_{i}, i\in e\}$ are well separated  when $\{v_{i}, i\in e\}$ are well separated.  Therefore, under the \MName{} model node sets with high \textit{diversity} are more likely to form hyperedges, i.e.,  nodes with latent position vectors as orthogonal as possible. In Supplemental Material F, we use simple \MName{} models to further illustrate this point, especially for the case when $\vert e\vert > d$. The probability $P(E=e)$ also depends on the lengths of $\widetilde{v}_{i}$'s ($i\in e$).  Since $\|\widetilde{v}_{i}\|_{2}^2=\|v_{i}\|_{2}^{2}+\alpha_{i}$ and $\|v_{i}\|_{2}>0$ is a constant as $i$ varies, $\|\widetilde{v}_{i}\|_{2}$ depends solely on $\alpha_{i}$. This means nodes with large  $\alpha_{i}$ values are more likely to form hyperedges.   Overall, hyperedge generation is driven by both within-hyperedge diversity (with respect to $v_{i}$'s) and nodes' popularity (represented by $\alpha_{i}$'s). 

\begin{remark} (Modeling $\mathcal{P}$ versus modeling Bernoulli distributions for each $e\subset[n_v]$). One important distinction between the proposed \MName{} model and most existing hypergraph models is that while most existing models assume independent Bernoulli distributions for whether there exists a hyperedge with configuration $e$ among all $e\subset[n_v]$ \citep{stasi2014beta,ghoshdastidar2017consistency,zhen2021community,ke2019community,lyu2021latent}, the \MName{} model considers hyperedges as i.i.d. realizations of one hyperedge distribution $\mathcal{P}$.  The structure of the \MName{} model is thus very different from that of previous works. The \MName{} model is especially suitable for scenarios where hyperedges form as `incidences', e.g. collaborations, purchase transactions, or assignments of tags to items. In these scenarios, the incidences of hyperedge formation can be reasonably assumed as independent, and any particular set of nodes can form hyperedges repeatedly. In contrast, the previous models are not very appropriate for this type of scenarios, since whether $e$ is observed as a hyperedge (which they model) largely depends on the number of observed hyperedges. 
It is also worth noting that while the likelihood functions of the previous models involve terms for all subsets $e$ of $[n_v]$ ($2^{n_v}$ of them in total), that of the \MName{} model only involves those that have ever been observed, which is usually much fewer than $2^{n_v}$. Therefore, the proposed modeling framework has a simpler likelihood function that allows the application of classical statistical inference tools, e.g., maximum likelihood estimation. \end{remark}

\begin{remark}[Choosing the way to incorporate the popularity parameters $\alpha$]
The DiPH model incorporates the popularity parameters $\alpha$ into $L$ additively, as in (\ref{definition::L}). Another potential way to include $\alpha$ is to let $L_{n_v\times n_v}=(\sqrt{\alpha_i\alpha_j}v_i^\top v_j)_{i,j=1}^{n_v}+I$,  which incorporates $\alpha$ multiplicatively. %
We deliberately choose to include $\alpha$ additively in order to decouple the effect of latent positions $v_i$'s and popularity parameters $\alpha$.  In Supplemental Material B, we compare the two formulations of $L$ and explain how (\ref{definition::L}) allows $\alpha$ to affect nodes' popularity more directly,  without incurring unnatural constraints to the model.
\end{remark}

\begin{remark}[Probability of $E=\{i,i'\}$]  The modeling of diversity and popularity in \MName{} can be further illustrated by calculating $P(E=\{i,i'\})$ for some $i,i'\in[n_v]$. Recall that $\|v_i\|_{2}^{2}$ is a constant across $i$, which we denote here by $\beta$. According to (\ref{definition::model}), we have
\begin{equation*}
P(E=\{i,i'\}) =\frac{L_{ii}L_{i'i'}-L_{ii'}^{2}}{\det(L+I)}=\frac{(\alpha_{i}+\beta)(\alpha_{i'}+\beta)-\cos^{2}({v}_{i},{v}_{i'})\beta^2}{\det(L+I)},
\end{equation*}%
where $\cos({v}_{i},{v}_{i'})$ denotes the cosine of the angle between vectors ${v}_{i},{v}_{i'}$. Since $\beta$ and the denominator are constants as $i$ and $i'$ vary, it is not difficult to see that $P(E=\{i,i'\})$ is large for nodes with large popularity parameters $\alpha_{i}$ and $\alpha_{i'}$ and with latent positions that are as orthogonal as possible (i.e. $\cos^{2}({v}_{i},{v}_{i'})$ being small). Also note that the `non-diversity' of the set $\{i,i'\}$ is quantified by $\cos^{2}({v}_{i},{v}_{i'})$, rather than $\cos({v}_{i},{v}_{i'})$. Therefore, if $v_{i}=-v_{i'}$, $\{i,i'\}$ is considered as non-diverse. \end{remark}

\begin{remark}[Probability of $E=\varnothing$] Under the \MName{} model, $P(E=\varnothing)$ is equal to $1/\det(L+I)$. While one may argue that $P(E=\varnothing)>0$ does not comply with real-world scenarios, we allow this minor misspecification to keep the model in a simple form, which also admits clarity in deducing marginal and conditional distributions under the model (see Section \ref{section::properties}). If one wishes to impose $P(E=\varnothing)=0$, (\ref{definition::model}) can be replaced by 
\[
P(E=e)=\frac{\det(L_e)}{\det(L+I)-1} ~~\text{for all}~e\neq\varnothing.
\]
\end{remark}

\subsection{Properties of the \MName{} model}

\label{section::properties}

The hyperedge distribution $\mathcal{P}$ in the proposed \MName{} model is in fact a specially structured (discrete) determinantal point process (DPP) \citep{kulesza2012determinantal}. A DPP is a type of distribution over the power set of a point set (e.g. $[n_v]$) as defined in (\ref{definition::model}), and the term $L$ in (\ref{definition::model}) for a generic DPP can be any positive semi-definite matrix. In the proposed \MName{} model, we require $L$ to take the special form as in (\ref{definition::L}). In addition to having a simple formulation, DPPs enjoy many nice mathematical properties and have received much attention in recent literature in applied mathematics and machine learning \citep{kulesza2012determinantal,BMRU2017,wilhelm2018practical,gartrell2019learning}. As a specially structured DPP, the \MName{} model inherits these properties. %
We discuss some of the important inherited properties of the \MName{} model here, which  
include the explicit and simple forms of marginal and conditional distributions, as well as the ease of sampling from DPP.%

\noindent \textbf{\emph{Marginal distribution.}} For any given $e\subset[n_v]$, $P(e\subset E)=\det(K_{e})$, where $K=I-(L+I)^{-1}$. In particular, for any $i\in[n_v]$, we have 
\begin{equation}
P(i\in E)=K_{ii}.\label{equation::margin}
\end{equation}
In Supplemental Material (33) and the discussion therein,
we characterize the relationship between $K_{ii}$ and node $i$'s the popularity parameter $\alpha_i$.

\noindent \textbf{\emph{Conditional distribution.}} For any $e_{1} \subset e_{2} \subset[n_v]$, we have 
\begin{equation}\label{equation::condition}
P(E=e_{2} \vert e_{1} \subset E) = \frac{\det(L_{e_2})}{\det(L+I-I_{e_1})},
\end{equation}
where we abuse the notation $I_{e_1}$ to represent a $n_v$-by-$n_v$ diagonal matrix whose diagonal entry is one if it is indexed in $e_1$ and is zero otherwise. In addition, as has been shown by \cite{borodin2005eynard}, conditioning on $e_1 \subset E$, $E \cap e_1^{c}$ is a DPP over the power set of $e_1^{c}$, the complement of $e_1$ in $[n_v]$. 

More properties of the \MName{} model are provided in the Supplemental Material (also see \citet{kulesza2012determinantal}), including how model parameters determine the distribution of $\vert E \vert$. For proofs of these properties, we refer the readers to \citet{kulesza2012determinantal}. Moreover, Algorithm 1 of \citet{kulesza2012determinantal} provides details for generating random samples from a given DPP, which also applies to the proposed \MName{} model. %

\begin{remark}[Background of DPP]\label{rem::DPP} The DPP is a type of distributions over the power set of a point (node) set, whose probability mass function is characterized by determinants. The DPP was firstly defined in the theory of quantum physics to model repulsion of particles in similar states \citep{macchi1975coincidence} and was also discovered in the distributions of eigenvalues of random matrices \citep{MEHTA1960420}. In the field of machine learning, it has been used for selecting a set of items while preventing redundancy, for example, in text summarization \citep{kulesza2011learning} and in recommendation systems \citep{wilhelm2018practical,chen2018fast,pmlr-v108-han20b}, where diverse sentences or products are selected.

More specifically, a (discrete) DPP is a distribution defined as in (\ref{definition::model}), where $L$ is any positive semi-definite matrix, which is referred to as its kernel matrix. \citet{kulesza2012determinantal} provide a detailed tutorial on the definition, properties and applications of DPPs. In addition to discrete DPPs, where the state space is a finite set, there are also studies on continuous DPPs, where the state space is (a region in) the Euclidean space, see e.g., \citet{lavancier2015determinantal,ghosh2020gaussian}.
\end{remark}

\begin{remark}[Comparison between the \MName{} model and other DPPs with compact parameterization] 
As parameterized by the kernel
matrix $L$, DPPs in general have $n_v(n_v+1)/2$ model parameters. By requiring $L$ to be formulated as in (\ref{definition::L}), the \MName{} model reduces the number of parameters to $n_vd+1$. It should be noted that the \MName{} model is not the only model that achieves a compact parameterization of DPPs by constraining structures on $L$. Previous works that achieve compact parameterization include \citet{gartrell2017low} and \citet{gartrell2019learning}, both of which assume certain low-rank structures of $L$. In contrast to the structure (\ref{definition::L}) in the \MName{} model, the low-rank structures in these previous models have the side effect of inducing the constraint that the generated point (node) set has a cardinality no larger than the rank, which is often unrealistic and too restrictive for real-world scenarios. \end{remark}

\subsection{Model parameter estimation}
\label{section::algorithm}

To fit the \MName{} model, we apply the maximum likelihood estimation (MLE) to an observed hypergraph with $\cE=\{e_1,e_2,\ldots,e_{n_e}\}$. Specifically, we estimate $v_i$'s and $\alpha_i$'s ($i\in[n_v]$) by solving 
\[
\underset{\substack{v_i\in\bR^{d},\alpha_i>0,\\ \|v_i\|_2 ~\text{is a constant over}~i}}{\arg\max} -\log\det(L+I) + \frac{1}{n_e}\sum_{\ell=1}^{n_e} \log\det(L_{e_{\ell}}),
\]
where $L_{n_v \times n_v} = (v_i^\top v_j)_{i,j=1}^{n_v}+\diag(\alpha)$.
To optimize the above criterion, we reparameterize $v_i$'s, $i\in[n_v]$
using $V_{n_v\times d}$ and $\beta$ by setting 
\begin{equation}
V_{i.} := \frac{v_i}{\|v_i\|_2} ~\text{for all}~ i, ~\text{and}~ \beta := \|v_i\|_2^2. \label{definition::3}
\end{equation}
Then the optimization problem above becomes 
\begin{equation}
\underset{\substack{V_{n_v\times d}, \|V_{i.}\|_2=1,\\
\beta>0,\alpha_i>0}}{\arg\max} -\log\det\left(\beta VV^\top + \diag(\alpha) + I\right) + \frac{1}{n_e}\sum_{\ell=1}^{n_e} \log\det \left( \beta(VV^\top)_{e_{\ell}} + \diag(\alpha)_{e_{\ell}} \right), \label{optimization::mle1}
\end{equation}
which allows a standard application of the \emph{projected gradient descent (ascent) algorithm} \citep{lange2013convex}. The algorithm iterates between the step of gradient update for parameters $V$, $\beta$, and $\alpha$ and the step of projecting the rows of $V$ into the $(d-1)$-dimensional unit sphere $\mathbb{S}^{d-1}$ and projecting $\beta$, $\alpha_1,\ldots,\alpha_{n_v}$
into $\bR^{+}$, the set of positive real numbers. After obtaining $\hat{V}$, $\hat{\beta}$ and $\hat{\alpha}$, we recover the estimate of $v_i$'s by calculating $\hat{v}_i=\hat{\beta}\hat{V}_{i.}$.

To accelerate the traditional projected gradient descent algorithm,  we apply the accelerated proximal gradient method developed by \citet{li2015accelerated} to solve (\ref{optimization::mle1}), which is particularly suitable for nonconvex objective functions. In addition, mini batch gradient update is utilized. Details of our algorithm are provided in Supplemental Material G,  where we also describe the approach of using Akaike information criterion (AIC) to choose the dimension $d$ of latent positions. 
We have implemented the algorithm in both R and Python, and the code is available upon request from the authors.

\section{Theoretical results}

This section establishes the consistency and asymptotic normality for the MLE estimates of the \MName{} model parameters. Note that owing to the constraints $\alpha_i>0$ and that $v_i$ has constant length across $i$ ($i \in [n_v]$), the parameter space is not a Euclidean space, but rather a specially structured manifold. This fact brings a major challenge to the theoretical development regarding the \MName{} model. Concerning the asymptotics of constrained M-estimation, we refer to the classical work of \citet{geyer1994asymptotics}.  
However, we note that when establishing the asymptotic normality for our model parameter estimates, extensive and non-trivial analysis on the local geometry of the parameter space is required. 

Consider a \MName{} model $H(v_1^*,\cdots,v_{n_v}^*,\alpha^*)$ with $v_{i}^{*}\in\bR^{d}$ and a hypergraph generated from it with hyperedges $\cE = \{e_{1}, \ldots, e_{n_e}\}$. The unknown model parameters $v_1^*,\cdots,v_{n_v}^*,\alpha^*$ are estimated using MLE. We denote the estimates obtained by solving (\ref{optimization::mle1}) as $\hat{v}_i$'s ($i\in [n_v]$) and $\hat{\alpha}$ and establish their consistency under the $l_2$ norm. Moreover, we also consider the $H(L^*)$ parameterization of the \MName{} model, where $L_{n_v\times n_v}^* = (v_i^{*\top}v_j^*)_{i,j=1}^{n_v} + \diag(\alpha^*)$.  The MLE for $L^*$ can be similarly defined as 
\begin{equation}
\hat{L} := \underset{L\in\cL}{\arg\max} -\log\det(L+I) + \frac{1}{n_e}\sum_{\ell=1}^{n_e} \log\det(L_{e_{\ell}}), 
\label{optimization::mle}
\end{equation}
where $\cL$ is the feasible space of $L$ in the \MName{} model. Specifically, we have 
\begin{equation}
\begin{split}
\cL & =\left\{ (v_i^\top v_j)_{i,j=1}^{n_v} + \diag(\alpha) \vert \alpha_i\in\bR^{+}, v_i\in\bR^{d}, \|v_i\|_2>0 ~\text{is constant across}~i, i\in[n_v] \right\} \\
& =\{\beta\cV+\text{diag}(\alpha) \vert \beta>0, \alpha_i>0 ~\text{for}~ i\in[n_v], \text{rank}(\cV)\leq d, \cV\succeq0, \text{diag}(\cV)=\bo\}.
\end{split}
\label{definition::Omega_L}
\end{equation}
Note that owing to the equivalence between the two parameterizations,
$\hat{L}$ can also be obtained by solving (\ref{optimization::mle1}) and setting $\hat{L}=\hat{\beta}\hat{V}\hat{V}^\top+\diag(\hat{\alpha})$. We will give the consistency of $\hat{L}$ under the Frobenius norm, and further, the asymptotic normality will also be established for $\hat{L}$.

\subsection{Consistency}

Recall that under the \MName{} model, $v_{i}^*$'s ($i\in[n_v]$) are identifiable up to a universal orthogonal transformation and individual sign flips, and $L^*$ is identifiable up to row and column sign flips (Theorem \ref{Thm::identifiability}). Therefore, we measure the error of $\hat{v}_{i}$'s ($i\in[n_v]$) using ${\min}_{O\in\mathbb{O}_{d}} \sum_{i=1}^{n_v} {\min}_{s=\pm1} \|\hat{v}_{i}-sOv_{i}^{*}\|_{2}$ and measure the error of $\hat{L}$ using ${\min}_{S\in\mathbb{D}_{n_v}} \|\hat{L}-SL^{*}S\|_{F}$, where $\mathbb{O}_{d}$ is the set of all $d$-by-$d$ orthogonal matrices and $\mathbb{D}_{n_v}$ is the set of all $n_v$-by-$n_v$ diagonal matrices with diagonal entries in $\{-1,1\}$. Theorem \ref{Thm::consistency_main} below establishes the consistency of the model parameter estimates.

\begin{theorem}\label{Thm::consistency_main} Given an observed hypergraph on $n_v$ nodes with $n_e$ hyperedges generated from the \MName{} model
$H(v_{1}^{*},\cdots,v_{n_v}^{*},\alpha^{*})$ with $v_i^*\in\bR^{d}$, consider the MLE estimates for the unknown parameters, which are $\hat{v}_{1},\cdots,\hat{v}_{n_v},\hat{\alpha}$, or $\hat{L}$ under the $H(L^{*})$ parameterization of the model. Assume that global optimums of (\ref{optimization::mle1}) are obtained for these estimates.  If $n_v>2d$ and $\{v_{1}^{*},\cdots,v_{n_v}^{*}\}$ span $\bR^{d}$, then as $n_e \rightarrow \infty$, we have 
\begin{align*}
\underset{O\in\mathbb{O}_{d}}{\min} \sum_{i=1}^{n_v} \underset{s=\pm1}{\min} \|\hat{v}_{i}-sOv_{i}^{*}\|_{2} & \overset{p}{\longrightarrow} 0,\\
\|\hat{\alpha}-\alpha^{*}\|_{2} & \overset{p}{\longrightarrow} 0,\\
\underset{S\in\mathbb{D}_{n_v}}{\min} \|\hat{L}-SL^{*}S\|_{F} & \overset{p}{\longrightarrow} 0.
\end{align*}
\end{theorem}

Note that the only conditions required to ensure the consistency of estimates are $n_v>2d$ and that $\{v_{1}^{*},\cdots,v_{n_v}^{*}\}$ span $\bR^{d}$, which are natural and easily satisfied.

Theorem \ref{Thm::consistency_main} is established using Theorem 5.14 in \citet{van2000asymptotic}, one of the classical theorems for establishing consistency of MLE estimates. The key step of applying that theorem is the construction of a compact subset of the parameter searching space to which the estimate almost surely belongs to. Though the technique for establishing the consistency in Theorem \ref{Thm::consistency_main} is relatively straightforward, that for the asymptotic normality in the next section is much less so.

\subsection{Asymptotic normality}

In this section, we develop the asymptotic normality for the MLE estimate of the \MName{} model parameters, considering the $H(L^{*})$ parameterization. We start by introducing some definitions, which will be used in the theorem.

\begin{definition}[Irreducibility of a matrix] Irreducibility is a property regarding square matrices, which we copy from \citet{BMRU2017}. We say that a given matrix $M_{n\times n}$ is reducible if there exists a partition $\{e_{1},e_{2},\cdots,e_{K}\}$ of $[n]$ with $K>1$, such that $M_{ii'}=0$ whenever $i\in e_{k}$, $i'\in e_{k'}$ and $k\neq k'$. Otherwise $M$ is said to be \textit{irreducible}.
\end{definition}

\begin{definition}[Vectorization of symmetric matrices] \label{definition::U} 
We define the function 
\[
\vecf : \bS_{n} \rightarrow \bR^{(n+1)n/2},
\]
where for any $M\in\bS_{n}$, $\vecf(M)$ is the vector formed by aligning the upper triangular part of $M$ by rows. Note that since $\vecf()$ is a bijection, $\vecf^{-1}()$ is naturally defined. Moreover, the Jacobian of $\vecf^{-1}()$ is a constant tensor, irrespective to where it is taken in $\bR^{(n+1)n/2}$, which we denote by $\bigtriangledown \vecf^{-1}$. In addition, for any $\mathcal{M}\subset \bS_{n}$, we define $\vecf(\mathcal{M}) := \{\vecf(M) \vert M\in \mathcal{M}\}$.
\end{definition}

\begin{definition}[Expected log-likelihood function] \label{definition::Phi}
We denote the expected log-likelihood function with respect to $L\in\cL$
as $\Phi(L)$. Specifically, we have 
\[
\Phi(L) = -\log\det(L+I) + \sum_{e\subset[n_v]} P_{L^{*}}(E=e) \log\det(L_{e}),
\]
where $P_{L^{*}}$ is the hyperedge distribution under $H(L^{*})$. Moreover, the Hessian of $\Phi$ at $L^{*}$ is denoted as $\bigtriangledown^{2}\Phi(L^{*})$. Note that since the argument of the function $\Phi$ is a matrix, $\bigtriangledown^{2}\Phi(L^{*})\in\mathbb{R}^{n_v\times n_v\times n_v\times n_v}$ is a fourth-order tensor. 
\end{definition}

\begin{definition}[Eigen decomposition of the matrix $(v_i^{*\top}v_j^*)_{i,j=1}^{n_v}$] \label{definition::eigen}
Given the $n_v$-by-$n_v$ matrix $(v_i^{*\top}v_{j}^*)_{i,j=1}^{n_v}$, $U^{*}$ and $U_{\perp}^{*}$ are a pair of matrices respectively having size $n_v\times d$ and $n_v\times(n_v-d)$, which together satisfy that $[U^{*},U_{\perp}^{*}]$ is orthogonal and 
\[
(v_i^{*\top}v_j^*)_{i,j=1}^{n_v} = [U^{*},U_{\perp}^{*}] \Lambda [U^{*},U_{\perp}^{*}]^\top,
\]
where $\Lambda$ is the diagonal matrix with diagonal entries being the eigenvalues of $(v_i^{*\top}v_j^*)_{i,j=1}^{n_v}$ in decreasing order. \end{definition}

\begin{definition}[Bouligand tangent cone of $\vecf(\cl)$ at $\vecf(L^{*})$] \label{definition::Bouligand}
Denote $\cl$ as the closure of $\cL$, where $\cL$ is defined in (\ref{definition::Omega_L}). We use $T_{\vecf(\cl)}(\vecf(L^{*}))$ to denote the Bouligand tangent cone of $\vecf(\cl)$ at $\vecf(L^{*})$. The explicit characterization of the elements in this cone is given by (80) in the Supplemental Material, and we also refer readers to \cite{geyer1994asymptotics} for more details on the Bouligand tangent cone. 
\end{definition}

Next, we establish the asymptotic normality of $\hat{L}$ in Theorem \ref{Thm::asymptotic}.

\begin{theorem}\label{Thm::asymptotic} Given an observed hypergraph on $n_v$ nodes with $n_e$ hyperedges generated from the \MName{} model $H(L^{*})$ with $L_{n_v \times n_v}^{*} = (v_i^{*\top}v_j^*)_{i,j=1}^{n_v} + \diag(\alpha^{*})$, $v_i^*\in\bR^{d}$ and $\alpha_i^* >0$, consider the MLE estimate $\hat{L}$ of $L^{*}$, obtained by the global optimum of (\ref{optimization::mle1}). Moreover, let 
\[
\tilde{L} := \underset{M \in \{S\hat{L}S \vert S\in\mathbb{D}_{n_v}\}}{\arg\min} \|M-L^{*}\|_{F}.
\]
If $n_v>2d$, $\{v_1^*,\cdots,v_{n_v}^*\}$ span $\bR^{d}$, and $(v_i^{*\top}v_j^*)_{i,j=1}^{n_v}$ is irreducible, then as $n_e\rightarrow\infty$, we have 
\begin{equation}
\sqrt{n_e} \cdot \vecf(\tilde{L}-L^{*}) \overset{dist.}{\longrightarrow} N\left({\bf 0}, -Q \left[ \bigtriangledown^{2} \Phi(L^{*})(\Gamma,\Gamma)\right]^{-1} Q^\top \right), 
\label{conclusion::normality}
\end{equation}
where $Q$ is an orthogonal matrix (i.e. $Q^{\top}Q=I$) with columns forming a basis for $T_{\vecf(\cl)}(\vecf(L^{*}))\subset \bR^{(n_v+1)n_v/2}$, $\Gamma$ is the result of multiplying $\bigtriangledown\vecf^{-1}$ with $Q$, $\bigtriangledown^{2}\Phi(L^{*})$ is the Hessian of $\Phi$ at $L^{*}$, and $\bigtriangledown^{2}\Phi(L^{*})(\Gamma,\Gamma)$ denotes the result of multiplying it with $\Gamma$ twice. 
\end{theorem}

Note that $\bigtriangledown \vecf^{-1}$ is a third-order tensor as $\vecf^{-1}$ maps a vector to a matrix. Accordingly, $\Gamma$ is also a third-order tensor. Recall from Definition \ref{definition::Phi} that $\bigtriangledown^{2}\Phi(L^{*})$ is a fourth-order tensor. When multiplying $\bigtriangledown^{2}\Phi(L^{*})$ with $\Gamma$, analogous to matrix product, slices (matrices) of the two tensors are taken and aligned according to the dimensions inherited from $L^*$, then inner products are calculated and the results are organized to form a third-order tensor. After performing this multiplication twice, we obtain $\bigtriangledown^{2}\Phi(L^{*})(\Gamma,\Gamma)$, which is a matrix (second-order tensor). Details of this multiplication are provided in the Supplemental Material, and we also refer readers to \cite{kolda2009tensor} for details on tensors and to \cite{BMRU2017} for the characterization of $\bigtriangledown^{2}\Phi(L^{*})$.

In comparison to Theorem \ref{Thm::consistency_main} (which shows consistency of the estimates), Theorem \ref{Thm::asymptotic} requires one more condition, which is that $(v_i^{*\top}v_j^*)_{i,j=1}^{n_v}$ is irreducible. This condition ensures that $\bigtriangledown^{2}\Phi(L^{*})(\Gamma,\Gamma)$ is invertible.

To the best of our knowledge, Theorem \ref{Thm::asymptotic} is the first result in the DPP literature that establishes the asymptotic normality for the parameter estimates of structured DPPs. Previous work by \citet{BMRU2017} proved the asymptotic normality for unstructured DPPs, where the parameter space of $L$ is the set of all positive definite matrices. That work is, indeed, one of the 
important building blocks when we establish Theorem \ref{Thm::asymptotic}. However, the major challenge for establishing Theorem \ref{Thm::asymptotic} stems from the fact that the \MName{} model specifies a much more structured parameter space (see (\ref{definition::Omega_L})) than the general DPP model studied in \citet{BMRU2017}. Our parameter space is far from linear, and is a highly structured manifold. We utilize Theorem 4.4 in \citet{geyer1994asymptotics} concerning constrained M-estimation to address this challenge and the proof of Theorem \ref{Thm::asymptotic} involves extensive examination of the local geometry of the manifold corresponding to the parameter space. In particular, classical as well as recent results in variational geometry \citep{rockafellar2009variational,BLPW2013,tam2017regularity} have been used to examine the manifold.

Theorem \ref{Thm::asymptotic} implies that $\|\Tilde{L} - L^*\|^2_F=[O_p(\sqrt{\tr (\Sigma)})+\tr (\Sigma)]/n_e$, where $\Sigma$ is the asymptotic covariance matrix in Theorem \ref{Thm::asymptotic}, i.e., $\Sigma := -Q [\bigtriangledown^{2} \Phi(L^{*})(\Gamma,\Gamma)]^{-1} Q^\top$. Therefore, as long as $n_e/\tr (\Sigma)\rightarrow \infty$, we have $\|\Tilde{L} - L^*\|_F=o_p(1)$. Note that $\Sigma$ is fully defined by the DiPH model parameters. It is also interesting to compare the result in Theorem \ref{Thm::asymptotic} with its counter part in \citet{BMRU2017} (Theorem 7 therein). The asymptotic covariance matrix given in \citet{BMRU2017} is $-(\bigtriangledown^{2}\Phi(L^{*})(\bigtriangledown \vecf^{-1},\bigtriangledown \vecf^{-1}))^{-1}$, which has full rank equaling $n_v(n_v+1)/2$. In contrast, the asymptotic covariance matrix under the \MName{} model, as characterized by $\bigtriangledown^{2}\Phi(L^{*})$ as well as $T_{\vecf(\cl)}(\vecf(L^{*}))$ in Theorem \ref{Thm::asymptotic}, has a largely reduced rank, which is no larger than $\rank(Q)\leq n_v(d+1)+1$, and this reduction is due to the structure (\ref{definition::L}) on $L$ that we impose.

\section{Simulation studies}%

In this section, we use simulation studies to investigate the effectiveness of the proposed algorithm (in Section \ref{section::algorithm}) and examine the performance of the MLE for the \MName{} model.  Simulation 1 examines the performance of MLE in estimating model parameters (i.e. $V,\beta,\alpha, L$) and simulation 2 demonstrates our method's effectiveness in node clustering. In simulation 2,  we include a method based on the degree-corrected hypergraph stochastic block model (DCHSBM) \citep{chodrow2021generative} for comparison,  as this is the only one we found in literature (see Table 1 in Supplemental Material) that handles the general hypergraph setting as we consider (i.e. hyperedges with multiplicity and varying cardinality). However,  despite the flexibility in model configurations considered by this method,  diversity is not supported by its algorithm.  We also include network community methods,  including the standard method of spectral clustering with normalization (NSC)  and the %
spectral clustering on ratios-of-eigenvectors (SCORE)
method \citep{jin2015fast,ke2023special}, which specializes on heterogeneous nodal popularity.  NSC and SCORE are applied on the dyadic projection of hypergraphs, as obtained by clique expansions.  Specifically, NSC with regularization (of size 0.1) is applied for its superior performance under sparse scenarios \citep{amini2013pseudo}; when calculating the coordinate-wise ratio in SCORE, instead of using the first eigenvector as the denominator, we used the row-wise $\ell_2$ norm of the trucated eigenvector matrix.

As in the previous section, we use the $l_{2}$-norm and the Frobenius norm as error measures to evaluate the performance of the estimates for the parameters of the \MName{} model $H(v_{1}^*,\cdots,v_{n_v}^*,\alpha^*)$ (or equivalently $H(L^*)$). Since the latent vectors have the same length, their directions (w.r.t. the origin) are especially important. In particular, we evaluate the direction and the length of $\hat{v}_i$ separately, which are denoted by $\hat{V}_{n_v\times d}$ and $\hat{\beta}$%
, where 
\[
\hat{V}_{i.} = \frac{\hat{v}_i}{\|\hat{v}_i\|_2} ~\text{and}~ \hat{\beta} = \|\hat{v}_i\|_2^2 ~\text{for all}~ i.
\]
Moreover, since $v_i^*$'s ($i\in[n_v]$) %
are only identifiable up to an orthogonal transformation and sign changes (Theorem \ref{Thm::identifiability}), we measure the error of %
$\hat{V}$ and $\hat{\beta}$ using 
\begin{equation}
\ell(\hat{V},V^*) := \underset{O\in\mathbb{O}_d,S\in\mathbb{D}_{n_v}}{\min} \|\hat{V}-SV^*O\|_{F} ~\text{and}~ \ell(\hat{\beta},\beta^*) := \vert\hat{\beta}-\beta^*\vert, 
\label{definition::lossV}
\end{equation}
where $V^*$ and $\beta^*$ are defined as in (\ref{definition::3}), and note that the rows of $V^*$ and $\hat{V}$ are all on the sphere $S^{d-1}$. The error of $\hat{\alpha}$ is evaluated using $\ell(\hat{\alpha},\alpha^*) := \|\hat{\alpha}-\alpha^*\|_{2}.$
When considering the $L$ parameterization of the \MName{} model, the error of $\hat{L}$ is evaluated using 
\begin{equation}
\ell(\hat{L},L^*) := \underset{S\in\mathbb{D}_{n_v}}{\min} \|\hat{L}-SL^*S\|_{F}.
\label{definition::lossL}
\end{equation}
Further, to aid interpretation, we report all the results in relative errors, including 
\[
\frac{\ell(\hat{V},V^*)}{\|V^*\|_{F}}, \frac{\ell(\hat{\beta},\beta^*)}{\vert\beta^*\vert}, \frac{\ell(\hat{\alpha},\alpha^*)}{\|\alpha^*\|_{2}} ~\text{and}~ \frac{\ell(\hat{L},L^*)}{\|L^*\|_{F}}.
\]

\begin{figure}[h!]
\centering \includegraphics[width=1\textwidth]{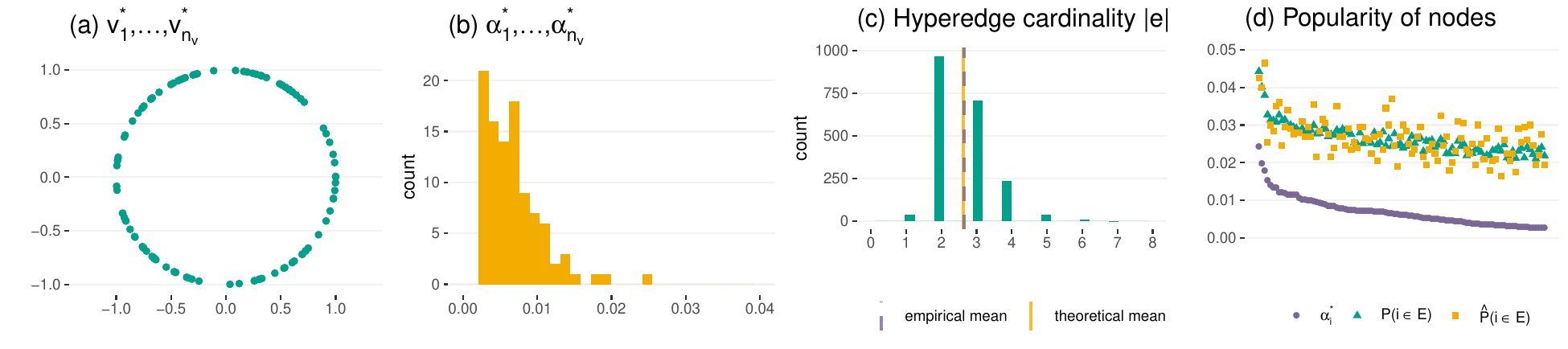}
\caption{Simulation 1 - an illustrative example ($d=2$, $n_v=100$ and $n_e=2000$). (a) The latent positions $v_i^*$'s are generated from the uniform distribution on $S^{d-1}$. (b) Histogram of the popularity parameters $\alpha_i^*$'s, which are generated using Beta distribution with transformations. (c) Histogram of $\vert e_{\ell}\vert$'s under (a) and (b), where $\mathbb{E}(\vert E\vert)$ is calculated using (23) in the Supplemental Material. (d) Theoretical and empirical node `popularities', where $P(i\in E)$ is calculated using (\ref{equation::margin}), and nodes are ordered by $\alpha_i^*$'s.
}\label{Fig::example} 
\end{figure}%
\subsection*{Simulation 1}
This is a setting where $v_1^*,\cdots,v_{n_v}^*$ are uniformly distributed on $S^{d-1}$. We consider \MName{} models with $n_v=100$ nodes and $d=2$, $3$ and $4$-dimensional latent positions respectively. The latent positions all have the unit length, and for a given $d\in\{2,3,4\}$, $v_1^*,\cdots,v_{n_v}^*$ are generated independently from the uniform distribution on $S^{d-1}$. To specify the popularity parameters, we first generate $\gamma_i \sim_{i.i.d.} \text{Beta}(1,4)$ and then set $\sqrt{\alpha_i^*}=0.15\gamma_i+0.05$ for $i\in[n_v]$. 
As a result, most nodes have a small popularity parameter value (with $\alpha_i^*$ close to $0.05^{2}$), while a few nodes have a relatively large popularity parameter value (with $\alpha_i^*$ approaching $0.2^{2}$). A hypergraph is then generated from $H(v_1^*,\cdots,v_{n_v}^*,\alpha^*)$ using Algorithm 1 of \citet{kulesza2012determinantal}, which simulates random samples for DPPs. Figure \ref{Fig::example} shows an illustrative example, together with summary statistics of a hypergraph (with $n_e=2000$). Note that in Figure \ref{Fig::example}(d), the empirical probability for node $i$ to be included in a hyperedge is calculated as follows: 
\begin{equation}
\hat{P}(i\in E) = \frac{1}{n_e} \sum_{\ell\in[n_e]} \mathbbm{1}(i\in e_{\ell}).
\label{definition::marg}
\end{equation}

\begin{figure}[h!]
\centering \includegraphics[width=1\textwidth]{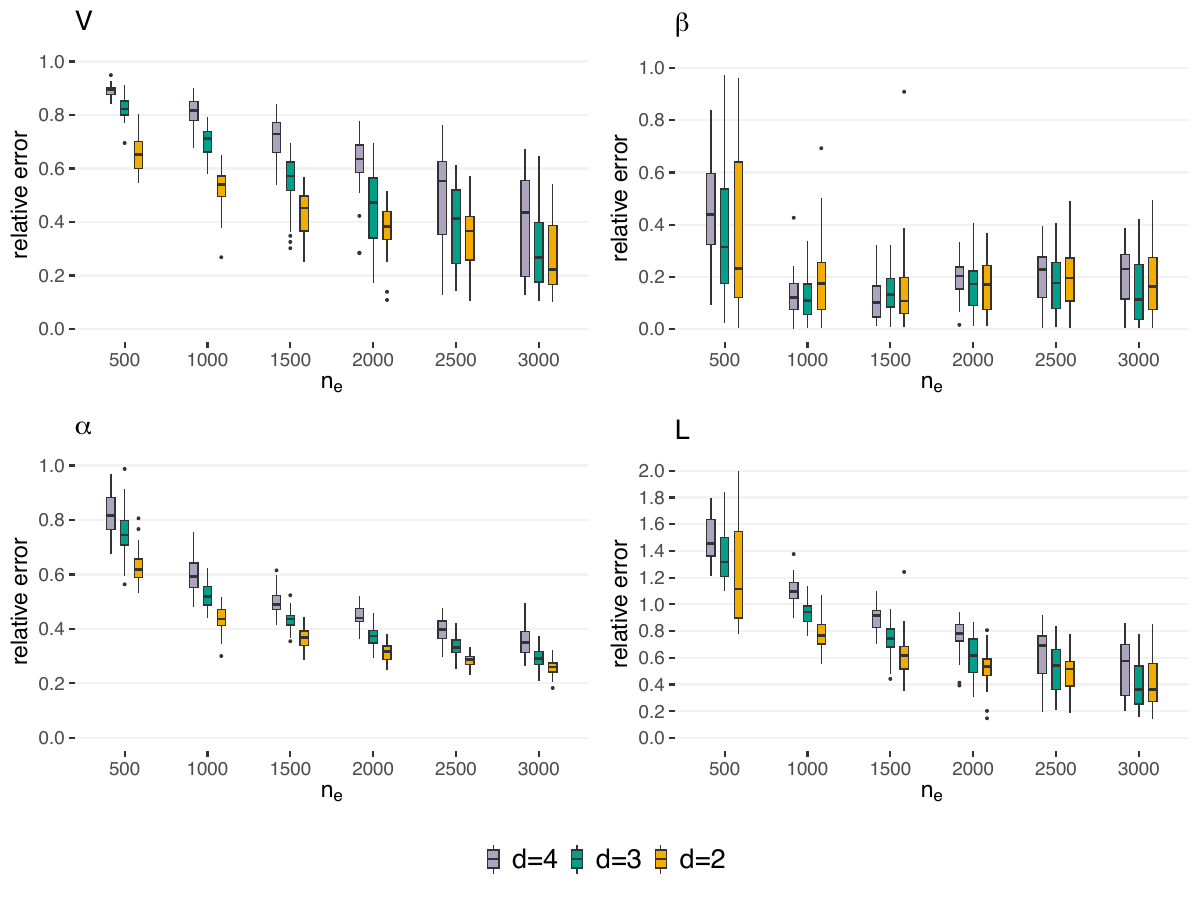} \caption{Simulation 1 - relative errors of $\hat{V}$, $\hat{\beta}$, $\hat{\alpha}$ and $\hat{L}$. Each box plot is based on results from 50 simulated hypergraphs. 
}
\label{Fig::consistency} 
\end{figure}

For each $d\in\{2,3,4\}$, we repeat the simulation 50 times. The number of hyperedges $n_e$ ranges from 500 to 3000. The results (in terms of relative errors) are reported in Figure \ref{Fig::consistency}. Note that when calculating $\ell(\hat{V},V^*)$ (\ref{definition::lossV}) and $\ell(\hat{L},L^*)$ (\ref{definition::lossL}), greedy searches for $S$ and $O$ were employed. %
Figure \ref{Fig::consistency} shows the effectiveness of the proposed algorithm in estimating the latent positions and the popularity parameters. It is also observed that as the number of hyperedges $n_e$ increases, the relative errors for all model parameter estimates decrease. Moreover, when the dimension of the latent positions $d$ increases, more hyperedges (i.e. larger $n_e$) are needed to obtain good quality estimates. 
The simulations here all use the true $d$ when fitting the \MName{} model. In Supplemental Material C.1, we show the effectiveness of selecting $d$ using the Akaike information criterion (AIC). Moreover, in Supplemental Material C.3 (Figure 8), we demonstrate Theorem \ref{Thm::asymptotic} by presenting the empirical distributions of the estimates from 500 repetitions.

\subsection*{Simulation 2}\label{sec::sim2}
\begin{figure}[h!]
\centering \includegraphics[width=\textwidth]{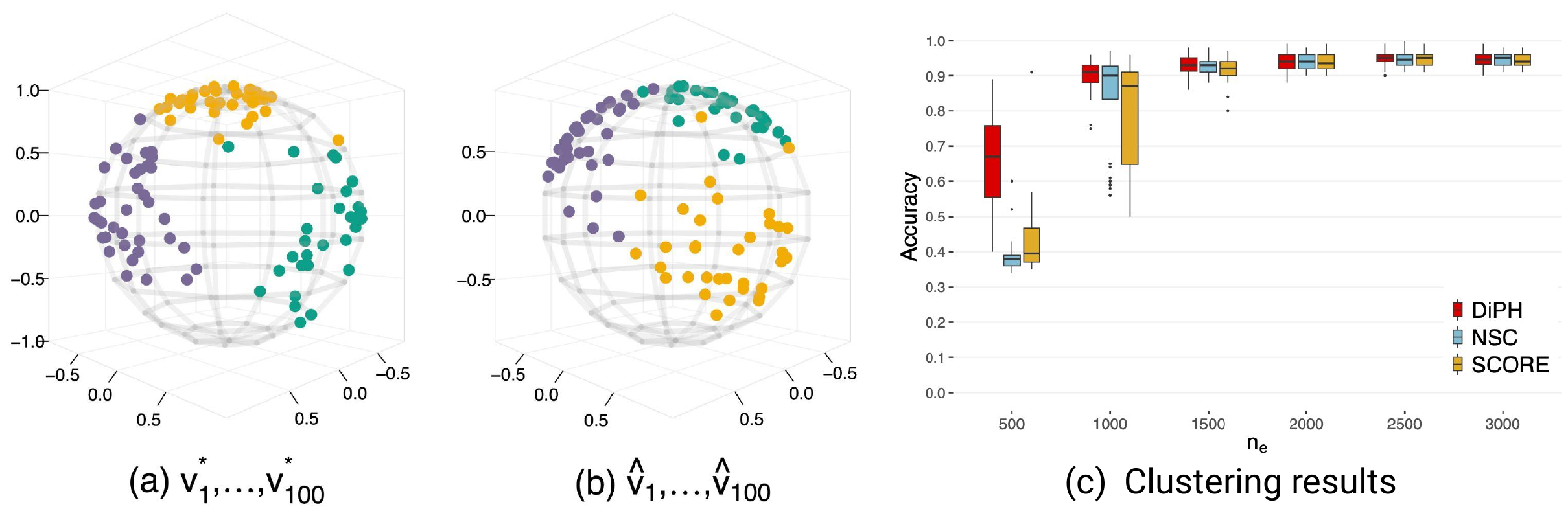}
\caption{Simulation 2 - (a) and (b) show an illustrative example where the nodes' latent positions are clustered. (a) The latent positions $v_i^*$'s are generated from three von Mise-Fisher distributions on $S^{2}$, which correspond to three clusters to which the nodes are uniformly assigned; nodes are colored according to cluster labels. (b) Estimates $\hat{v}_i$'s are obtained using the proposed algorithm; nodes are colored according to their true cluster labels.  (c) Clustering accuracy of applying line k-means to $\hat{v}_i$'s (denoted as DiPH),  and of NSC and SCORE,  as applied to the weighted network resulted from clique expansions of hypergraph; we set the number of clusters to three for all clustering methods. }%
\label{Fig::sim_cluster_sphere}
\end{figure}

This is a setting where $v_1^*,\cdots,v_{n_v}^*$ have a clustered structure. 
In this simulation study, we randomly assign each of the $n_v=100$ nodes to one of three clusters with equal probability. The latent positions of nodes $v_1^*,\cdots,v_{n_v}^*$ are then generated independently from von Mise-Fisher distributions parameterized by the concentration parameter $\kappa=10$ and the mean directions $\mu=(1,0,0)$, $(0,1,0)$ and $(0,0,1)$ respectively corresponding to the three clusters. Note that the von Mise-Fisher (vMF) distribution is the spherical analogue of the normal distribution on the unit sphere. Specifically, its probability density function is given by $f(v)=C(\kappa) \exp\left(\kappa\mu^{\top}v\right)$, where $C(\kappa)$ is the normalizing constant.  Figure \ref{Fig::sim_cluster_sphere}(a) shows an example of nodes' latent positions generated as described above, where nodes are colored according to their assigned cluster labels. Other than the nodes' latent positions, we conduct the simulation study in the same way as in the previous simulation study, using the same procedure to generate $\alpha$, the same values for $n_e$, and the same number of repetitions. Figure \ref{Fig::sim_cluster_sphere}(b) shows an example of $\hat{v}_1,\cdots,\hat{v}_{n_v}$ when $d=3$, $n_v=100$, and $n_e=2000$.  In Figure 9 (Supplemental Material C), we see that when $n_e$ increases, the relative errors of $\hat{V}$, $\hat{\beta}$, $\hat{\alpha}$ and $\hat{L}$ all decrease, a trend similar to that in the previous simulation study. In addition to parameter estimation errors, here we also examine if the fitted \MName{} model can recover the cluster structure of nodes. Recall that $\hat{v}_i$'s have uniform lengths and are free to sign flips; or more precisely, $\hat{v}_i$ and $-\hat{v}_i$ are equivalent.  We develop a k-means algorithm that addresses these special characteristics of $\hat{v}_i$'s, which we call the \textit{line k-means algorithm}. Details are provided in Supplemental Material G.4.  Figure \ref{Fig::sim_cluster_sphere}(c) shows the clustering accuracy of this method, together with that of network community detection methods NSC and SCORE.  It can be seen that the proposed method largely outperforms network community detection methods when the number of hyperedges $n_e$ is relatively small.  As $n_e$ increases, the clustering accuracy of all methods increases,  and they gradually become similar.  These results demonstrate the effectiveness of using $\hat{v}_1,\cdots,\hat{v}_{n_v}$ to recover nodes' cluster labels. 

We also include two more simulations in Supplemental Material C.2,  including settings where the number of clusters is seven and where the number of nodes $n_v$ is 500.  The corresponding results (Figure 7) further demonstrate the effectiveness of the proposed DiPH based hypergraph community detection method in various settings.  Moreover, we provide additional results for Simulation 2 in Supplemental Material C,  including the comparison with another hypergraph community detection method that is based on DCHSBM.  We also investigate the effectiveness of using multiple model parameter initializations to improve the accuracy of fitting the DiPH model (Table 3). Reports of running time of all compared methods are provided in Supplemental Material as well.  Additionally,  we compare using line k-means versus fitting a mixture of von Mises-Fisher distributions as the last step of the DiPH based community detection method (Figure 11) and show the superiority of the proposed line k-means algorithm.

\section{A cooking recipe data example}\label{sec::cook}

In this section, we apply the proposed \MName{} model to analyze the \textit{What's Cooking} data \citep{cookData}. The data set comprises 39,774 recipes on yummly.com, where each recipe contains a list of ingredient names. Naturally, each recipe can be considered as a hyperedge of ingredients, and for our analysis, we focus on all the recipes in the Chinese cuisine, which consist of 2,673 recipes with 1,792 ingredient names. We preprocessed the data so that ingredient names of the same or very similar meanings were combined (e.g. `soy sauce' and `regular soy sauce', `green onion' and `spring onion'), and ingredients that appeared in only one recipe were deleted. After preprocessing, we obtained a total number of 906 ingredients. Viewing each recipe as a hyperedge of ingredients and the collection of all Chinese cuisine recipes as a hypergraph ($n_v=906$, $n_e=2673$), we fitted the \MName{} model, and to enable easy visualization, we chose $d=3$. 

\begin{figure}[h!]
\centering \includegraphics[width=1\textwidth]{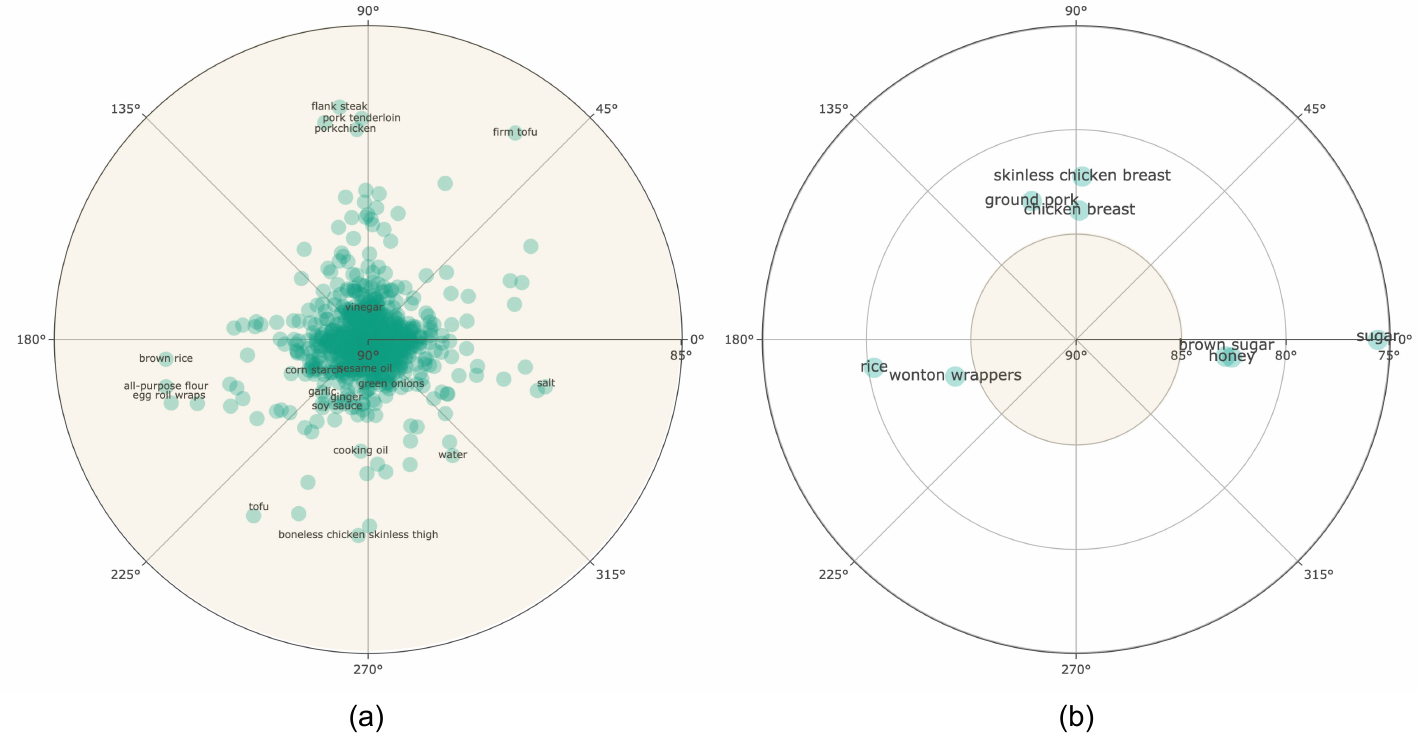}
\caption{Results from the estimated \MName{} model for Chinese recipes in the \textit{What's Cooking} data. (a) and (b) show the estimated latent positions in the spherical coordinate system, with longitude and latitude as the axes, where (a) shows all estimated latent positions at latitude larger than 85$^{\circ}$ and (b) is a zoom-out view only showing estimated latent positions with latitude smaller than 85$^{\circ}$.  The ten most frequently used ingredients as well as the ten ingredients furthest from the `north pole' in (a) are annotated.  %
}
\label{Fig::recipe} 
\end{figure}

Figure \ref{Fig::recipe}(a) and \ref{Fig::recipe}(b) show the estimated $\hat{v}_{i}$'s in the spherical coordinate system (with longitude and latitude) where the center is the `north pole', (a) corresponds to the region with latitude larger than 85$^{\circ}$, and (b) is a zoom-out view but only showing embedded ingredients with latitudes smaller than 85$^{\circ}$ (note that the inner colored region in (b) corresponds to (a)). The ten most frequently used ingredients as well as the ten ingredients furthest from the `north pole' in (a) are annotated. It is observed that most ingredients concentrate in a high latitude region (Figure \ref{Fig::recipe}(a)) while a few ingredients spread outward in roughly three directions - up, left and right (Figure \ref{Fig::recipe}(b)). It is also observed that the direction around the 90$^{\circ}$ longitude comprises many proteins, such as various meats, and the direction with longitude between 180$^{\circ}$ and 225$^{\circ}$ is mainly composed of carbohydrate ingredients, including both rice and wheat products. Note that if a similarity driven model is used, it is unlikely to see meat (carbohydrate) ingredients cluster together, as a Chinese dish rarely uses more than one type of meats (carbohydrates). For example, \citet{ahn2011flavor} found that East Asian cuisines tend to avoid compound sharing ingredients. On the other hand, by exploiting diversity, the \MName{} model is able to identify the meat and carbohydrate categories, the two most commonly used staple foods in Chinese cuisine. It is also observed that the very high latitude region around the `north pole' comprises mainly seasonings, such as vinegar, green onions, ginger, garlic, etc. Figure 12 (in Supplemental Material) shows the estimated popularity parameters $\hat{\alpha}_i$'s, which align well with the frequencies that ingredients appear in recipes. %

\begin{figure}[h!]
\centering \includegraphics[width=\textwidth]{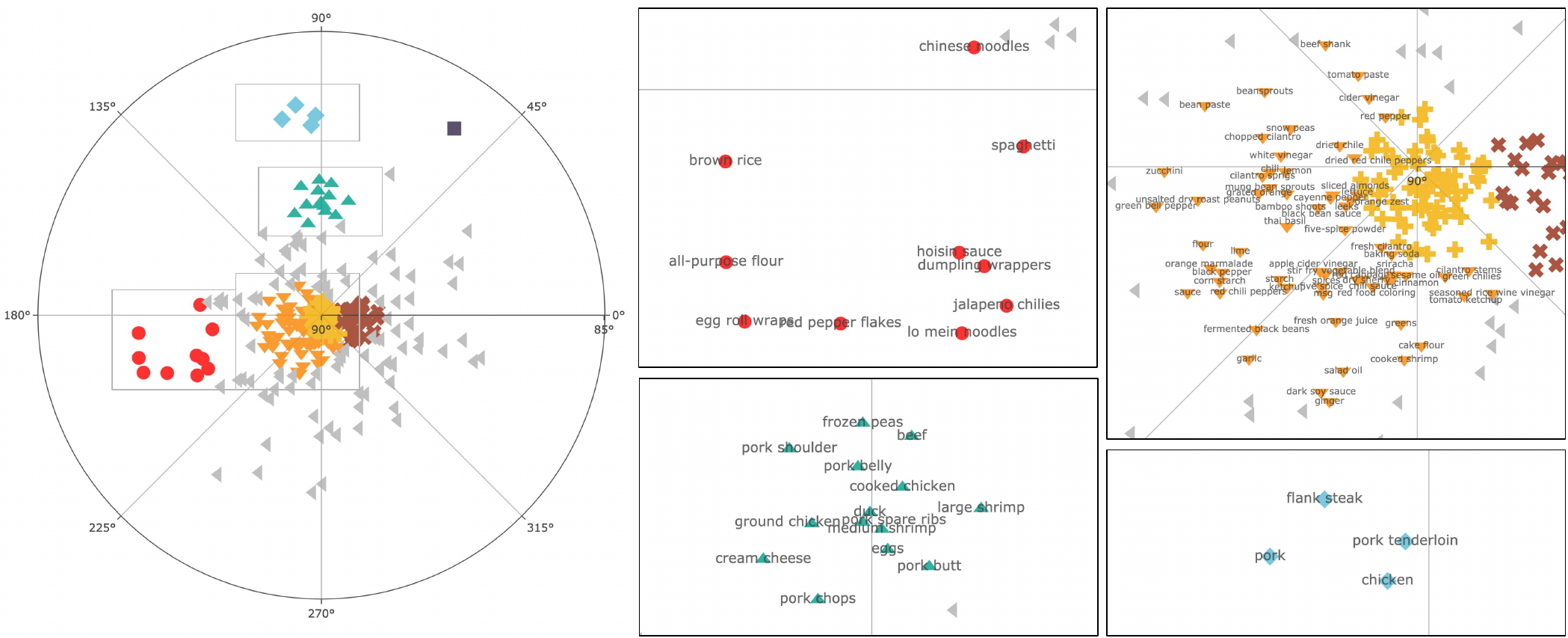}
\caption{Clustering of ingredients based on estimated latent positions. The 298 ingredients that have appeared at least 10 times in recipes are clustered into 8 groups (denoted using 8 different point shapes) by fitting a mixture of von Mise-Fisher distributions. 
Enlarged views of 4 clusters are shown on the right hand side, which comprise mainly carbohydrates, seasonings and proteins respectively.}
\label{Fig::recovery_cool} 
\end{figure}

The fitted \MName{} model has various utilities, including generating and completing hyperedges. For example, the explicit formula of the conditional probability $P(E=e_2\vert e_1\subset E)$ (in Section \ref{section::properties}) can be used to complete a hyperedge that is known to comprise elements of $e_1$. In addition, the estimated latent positions and popularity parameters can be used for various downstream tasks, e.g. clustering and supervised learning. Here, we demonstrate the clustering application using the fitted model of recipes. The demonstration is done on the 298 ingredients that have appeared 10 or more times in recipes, as most of the ingredients that appear fewer than 10 times concentrate around the `north pole'. %
We fit a mixture of von Mise-Fisher distributions to the estimated latent positions of the 298 ingredients. The Akaike's Information Criterion (AIC) and Bayesian Information Criterion (BIC) respectively select 8 and 3 as the number of clusters, and we show the result of 8 clusters in Figure \ref{Fig::recovery_cool}. 
Notably, there is one cluster mainly comprising carbohydrates (circles), one for seasonings (downward pointed triangles), and two for proteins (diamonds and upward pointed triangles respectively). %
We refer the readers to the Supplemental Material for full details of the clustering result. %

\section{Conclusion}

In this paper, we proposed a new hypergraph latent space model, \MName{}, which allows hyperedges with varying cardinality.
More importantly, it is the first hypergraph model that is driven by diversity (rather than similarity) of nodes within hyperedges, which is commonly seen in real-world hypergraphs. In addition, the proposed model admits heterogeneity in the popularity of nodes. We have established the consistency and asymptotic normality for the MLE estimates of the model parameters. \\
\indent Potential real-world applications of  the proposed model include team collaboration hypergraphs, where members with diverse expertise are preferred; grocery shopping cart hypergraphs, where redundant items are avoided; and medical International Classification of Diseases (ICD) code hypergraphs in electronic health records. The fitted models can support various down stream analyses, including clustering (as illustrated in Section \ref{sec::cook}) and supervised learning, where we treat $\hat{v_i}$ as latent features of nodes. The fitted \MName{} model can also be used to complete a partially observed hyperedge that is known to comprise elements of $e\subset [n_v]$, for example, by solving $\arg\max_{i\in [n_v]/e}P(E=e\cup\{i\}\vert e\subset E)$ using (\ref{equation::condition}). Further, fitted \MName{} models can be used to generate random hyperedges, which can also be done with a desired hyperedge size (see Supplemental Material E).

The \MName{} model is a determinantal point process (DPP) with a special structure. This structure allows a significant reduction in the number of model parameters, while not introducing artificial upper bounds on hyperedge cardinalities. Moreover, in the literature concerning DPP, our theory is the first to our knowledge that establishes the asymptotic normality for structured discrete DPP.

\bigskip{}

\bibliographystyle{apalike}
\bibliography{bib-hypergraph}

\end{document}